# Emotional Expression Detection in Spoken Language Employing Machine Learning Algorithms

Mehrab Hosain[1], Most. Yeasmin Arafat[2], Gazi Zahirul Islam[3], Jia Uddin[4], Md. Mobarak Hossain[5], and Fatema Alam[6]

[1]Department of Computer Science and Engineering, University of Information Technology & Sciences, Dhaka-1212, Bangladesh
email: robinhosain@gmail.com

[2]Department of Computer Science and Engineering, University of Information Technology & Sciences, Dhaka-1212, Bangladesh
email: yearashefa@gmail.com

[3]Department of Computer Science and Engineering, Southeast University, Dhaka-1208, Bangladesh
email: gazi.islam@seu.edu.bd

[4]AI and Big Data Department, Endicott College, Woosong University, Daejeon, Korea,
email: jia.uddin@wsu.ac.kr

[5]Department of Computer Science and Engineering, Daffodil International University, Dhaka-1216, Bangladesh
mobarak15-9636@diu.edu.bd

[6]Department of Physics, Jahangirnagar University, Dhaka-1342, Bangladesh
email: fatemazinnah89@gmail.com

**Abstract:**

There are a variety of features of the human voice that can be classified as pitch, timbre, loudness, and vocal tone. It is observed in numerous incidents that human expresses their feelings using different vocal qualities when they are speaking. The primary objective of this research is to recognize different emotions of human beings such as anger, sadness, fear, neutrality, disgust, pleasant surprise, and happiness by using several MATLAB functions namely, spectral descriptors, periodicity, and harmonicity. To accomplish the work, we analyze the CREMA-D (Crowd-sourced Emotional Multimodal Actors Data) & TESS (Toronto Emotional Speech Set) datasets of human speech. The audio file contains data that have various characteristics (e.g., noisy, speedy, slow) thereby the efficiency of the ML (Machine Learning) models increases significantly. The EMD (Empirical Mode Decomposition) is utilized for the process of signal decomposition. Then, the features are extracted through the use of several techniques such as the MFCC, GTCC, spectral centroid, roll-off point, entropy, spread, flux, harmonic ratio, energy, skewness, flatness, and audio delta. The data is trained using some renowned ML models namely, Support Vector Machine, Neural Network, Ensemble, and KNN. The algorithms show an accuracy of 67.7%, 63.3%, 61.6%, and 59.0% respectively for the test data and 77.7%, 76.1%, 99.1%, and 61.2% for the training data. We have conducted experiments using Matlab and the result shows that our model is very prominent and flexible than existing similar works.

## 1. INTRODUCTION

The human voice is very complex and can show a wide range of emotions. Emotion in speech gives information about how people act or feel. There are many functions of the human vocal system that make it possible for humans to speak. These include tone, pitch, energy, entropy, and many other aspects of the speech. The increasing need for human-machine interactions indicates that more tasks to be done to improve the results of these interactions, like giving computer and machine interfaces the ability to understand how a person feels when they speak. Emotions play a big part in how people talk to each other. People and machines should be able to work together more effectively if computers have built-in skills for figuring out how people feel [2], [5]. Today, a lot of resources and time is being spent on improving artificial intelligence and smart machines to make our life easier and more comfortable. According to the findings of several pieces of literature, human feelings regulate the decision-making process to some extent [1]-[4]. If the machine can figure out how people are feeling when they speak, it will be able to respond and communicate properly.

Recognizing people's feelings based on what they say is still a challenge. The authors in [3] propose CNNs (Convolutional Neural Networks), RNNs (Recurrent Neural Networks), and time distributed CNNs based networks in their study, but they do not use any usual hand-crafted features that are typically used to identify emotional speech. To solve the SER (Speech Emotional Recognition) problem, they integrate LSTM (Long Short-Term Memory) network layers into a deep hierarchical CNN feature extraction architecture. They have got almost 87% and 78% accuracy.

## 2. RELATED WORKS

Previously, the detection of emotions in auditory signals is studied rigorously. Those published works [17]-[21] in this field use a variety of classifiers including the SVM, NN, CNN, LSTM, and Bayes Classifier. The number of emotions is varied in different studies and the researchers are very cautious in determining the accuracy of the classifiers. We observe that reducing the number of emotions for recognition provides very good accuracy. Table 1 summarizes some past research papers on this study.

Table 1: Review and comparison of some prominent articles

| Study | Feature Extraction | Algorithm | No. of Emotions | Accuracy |
|---|---|---|---|---|
| 1. "Speech based Emotion Recognition using Machine Learning" [2021] [2]. | MFCC | CNN | 6 | 88.21% |
| 2. "Speech Emotion Recognition with Convolutional Neural Network" [2019] [10]. | MFCC | CNN | 8 | 72% |
| 3. "Speech Emotion Recognition using Convolutional and Recurrent Neural Networks" [2016] [3]. | LSTM | CNN/RNN | 7 | 86.23/78.83% |
| 4. "Speech Emotion Recognition using SVM" [2020] [1]. | MFCC, LPCC | SVM | 4 | 85.085/73.25% |
| 5. "Speech Emotion Recognition Using Fourier Parameters" [2015] [5]. | FP- Furrier Parameter | SVM | 6 | 87.5 |
| 6. "Voice Based Emotion Recognition with Convolutional Neural Networks for Companion Robots" [2017] [6]. | MFCC & GMM | CNN | 6 | 71.33 |
| 7. "Research on speech emotion recognition based on deep auto-encoder" [2016] [7]. | Deep auto-encoder (DAE) | SVM | 6 | 86.4 |

Audio-Visual Emotion Recognition (AVER) [6] evaluates four human feelings (i.e., happiness, sorrow, rage, neutral) using a Bayesian classifier. The classifier approximates 80% for happiness and sorrow and 65% for rage. The efficiency for neutral feelings is not very good in the SVM fusion method while anger and melancholy are roughly estimated at 80% accuracy. The accuracy for happiness is not good enough which is 60% due to the poor articulation of the voice modality. The combination of effective multimodal stimuli to obtain high recognition accuracies for all classes is a difficult task.

Using DAE (Deep Auto Encoder) [7], five hidden layers were created. The researchers of [7] evaluate DAE to the standard feature techniques Mel-frequency cepstral coefficients (MFCCs), and Linear Predictive Cepstral Coefficients (LPCC). The finding indicates that six emotions have the highest recognition accuracy rate i.e., 86.41%. Earlier voice-based emotion identification systems concentrate mostly on speech collection.

We use SVM (Support Vector Machine), NN (Neural Network), KNN (K-Nearest Neighbours), and Ensemble classifiers and feature extraction techniques MFCC and GTCC. Table 1, summarizes some literature. We may consider the EMD (Empirical Mode Decompose) for feature decomposition and extraction. The rest of the paper explains the related works, methodology, dataset, and result.

## 3. METHODOLOGY

*3.1 Empirical Mode Decomposition*

Empirical mode decomposition, widely known as EMD, is a data-adaptive multiresolution approach that can be used to deconstruct a signal into components that have physical significance. In EMD, non-linear and non-stationary signals can be analyzed by first being broken down into their component parts using a variety of various resolutions. The domains of fault identification, biological data analysis, power signal analysis, and seismic signal analysis are some of the more common uses of empirical mode decomposition. The EMD is an adaptive manner into a finite set of AM/FM modulated components by utilizing a process which is known as Sifting Algorithm [15]. The components that represent the oscillation modes ingrained in the data, are referred to as Intrinsic Mode Functions or simply IMFs. An IMF is a function for which the number of extrema and the number of zero crossings differ by a maximum of one, and the mean of the two envelopes associated with the local maxima and local minima is approximately zero. A valid mathematical construct is required to express the IMF. A signal, *x(k)*, can be broken down into its component parts using the formula below,

$$x(k) = \sum_{i=1}^{M} c_i(k) + r(k) \qquad (1)$$

where $c_i(k)$, $i = 1, \dots, M$, is the set of IMFs and $r(k)$ is the residual. The first IMF is obtained as follows [1],

1. Let $\tilde{x}(k) = x(k)$;
2. Identify all local maxima and minima of $\tilde{x}(k)$;
3. Find an "envelope," $e_{min}(k)$ (resp. $e_{max}(k)$) that interpolates all local minima (resp. maxima);
4. Extract the "detail," $d(k) = x(k) - (1/2)(e_{min}(k) + e_{max}(k))$;
5. Let $\tilde{x}(k) = d(k)$ and go to step 2; repeat until $d(k)$ becomes an IMF.

   Once the first IMF is obtained, the procedure is applied to the residual, $r(k) = x(k) - d(k)$ to obtain the second IMF. Similarly, the procedure is applied recursively to obtain all the IMFs. An example of EMD is shown in Figure 1, which illustrates a segment of speech.

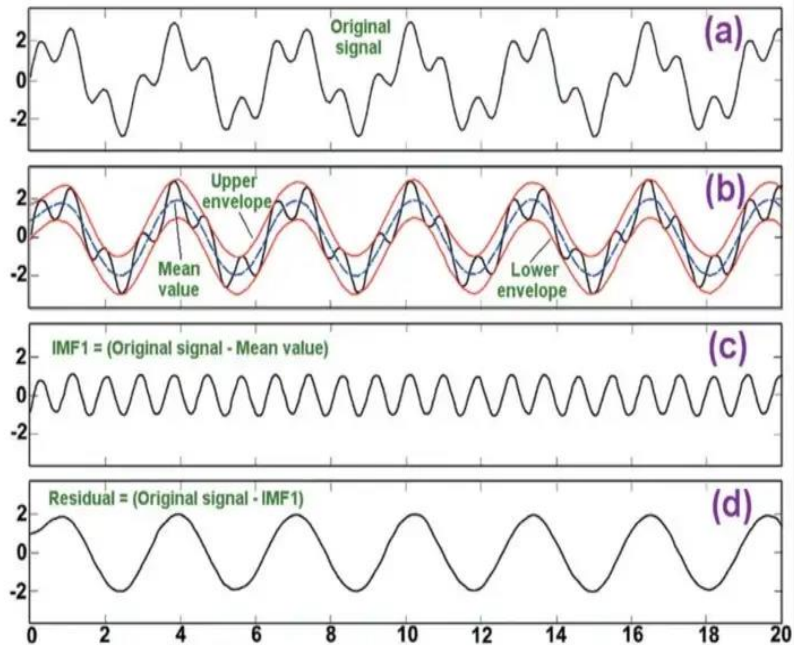

Figure 1: Signal analyzed with empirical mode decomposition

Time-frequency analysis can be conducted by leveraging the use of empirical mode decomposition, without leaving the time domain. Since the sub-signals share the same temporal scale as the original signal, dissection is simplified. Empirical mode decomposition is a Multi-Resolution Analysis (MRA) method that recursively extracts alternative resolutions from the data itself without the need for fixed functions or filters.

*3.2  Feature Extraction*

To train machine learning or deep learning systems, we extract features from audio sources and input those into classifiers. When designing a feature extraction pipeline, it is best to use functions, such as melSpectrogram, mfcc, pitch, spectralCentroid, and audioFeatureExtractor to reduce unnecessary calculations. Following the feature extraction process, we evaluate the mean and variance for each data set.

3.2.1 Auditory Spectrograms

The audioDelta function represents the local slope using least squares over an area centered on sample $x(k)$, which includes M samples before and after the current sample.

3.2.2 Auditory Cepstral Coefficients

**Gammatone cepstral coefficients (gtcc)**: Figure 2 shows the block diagram of gtcc. The calculation of the Gammatone Cepstral Coefficients (GTCC), also known as Gammatone Frequency Cepstrum Coefficient (GFCC) is similar to the MFCC extraction method. The audio stream is first windowed into brief frames, typically lasting between 10 and 50 ms. There are two goals for this approach to extract gammatone cepstral coefficients, delta, and log-energy.  The gtcc function divides the data into many overlapping parts. Overlap length specifies the duration of overlap between analysis windows.

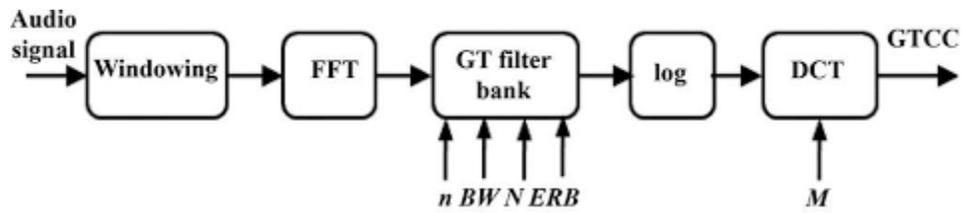

Figure 2: Block diagram of gtcc

The non-stationary audio signal can be deemed to be stationary for a short time. Following that, the Fast Fourier transform (FFT) of the signal is applied to the GT filter bank, which emphasizes the perceptual significant signal frequencies.

The design of the GT filter bank is the main purpose of this research, which considers several aspects such as the total filter bank bandwidth, the order of the GT filters, the ERB (Equal Rectangular Bandwith) model, and the number of filters. Finally, the discrete cosine transform (DCT) is used to replicate human loudness perception and decorrelate the outputs of the logarithmically compressed filter. This results in improved energy compaction. The total cost of computing is roughly the same as the MFCC computation.

**Mel-Frequency Cepstral Coefficients (mfcc)**: The block diagram of MFCC is shown in Figure 3. The feature extraction technique is a popular method for extracting speech features. Therefore, ongoing research is trying to improve its performance. One of the newer MFCC implementations (i.e., Delta-Delta MFCC) increases speaker verification.

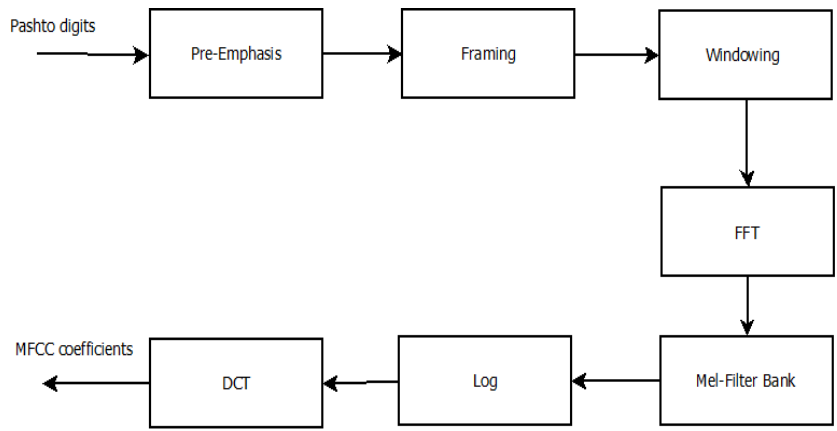

Figure 3: Block diagram of MFCC

3.2.3 Periodicity and Harmonicity

The harmonic ratio represents the connection between the fundamental frequency's strength and the overall strength of an audio sample. It's a measurement of the amount of harmony in a signal. Physical characteristics of sound that correlate to amplitude, time, and variables connected to frequency are known as pitch. Pitch is the physical feature of sound that corresponds to timbre, volume, and rhythm.

3.2.4 Spectral Descriptor

Machine and deep learning systems, as well as perceptual analysis, make extensive use of spectral descriptors. Spectral descriptors have been used in a variety of applications. It describes the shape that is sometimes referred to as the *timbre* of audio. These features can be used to characterize the spectral content of the audio signals. Timbre and coloration are two examples of these types of perceptual attributes. All of the spectral features are computed for individual time frames. Spectral features provide a summary of the spectral content of the rate-map representation across all of the auditory filters. Some of the spectral descriptor features are spectralCentroid, spectralEntropy, spectralFlux, spectralRolloffPoint, and spectralSpread.

## 3.3 Classification Algorithms

### 3.3.1 Support Vector Machine

SVM stands for support vector machine and is a supervised learning approach that is commonly utilized for pattern recognition applications [22], [24]-[26]. Even when trained on a small training dataset, the method still gives satisfactory results because it is so easy to use. To be more specific, it is an algorithm that generates a hyperplane or set of hyperplanes in a space that has high dimensions or an infinite number of dimensions. These hyperplanes have the potential to be utilized for regression and classification tasks. Its goal is to find a hyperplane that maximizes the separation between data points that belong to distinct classes, which will ultimately result in a more accurate classifier. There is a possibility that the data can be separated linearly or non-linearly. A kernel function is used to classify data that is non-linearly separable. This function turns the feature space of the data into a high-dimensional space. Because of this, it is possible to linearly separate the data points in this high-dimensional space. In this high-dimensional space, the data points are therefore linearly distinct from one another. The optimization challenge is simplification with the use of SVM.

$$a = min\left(\frac{\|w\|^2}{2}\right) \text{ subject to } \forall k, y_k(<w \cdot x_k> +b) \geq 1 \tag{2}$$

Where $w$ is the hyperplane's normal vector, $<w.x_k>$ denotes the value of the inner product of $w$. The $x_k$ for each data point in the training set is $(x_k, y_k)$. The radial bias kernel and the linear kernel are two frequently used kernels.

The formula for the linear kernel function is Kernel $(x, y) = (x, y)$. The radial bias kernel is expressed by the following equation,

$$kernel(x, y) = e^{\frac{-\|x-y\|^2}{(2\sigma^2)}} \tag{3}$$

### 3.3.2 Neural Network

The Neural Network is one of the most common ways to solve classification problems. The most significant trait of the NN is to use the training network to find the best settings for the network. The process of optimizing the network is based on reducing error functions like the cross-entropy error function shown in equation (4) [23]. But over-fitting can happen when the training data are more than enough. To keep from over-fitting, early stopping can be used to stop training when the validation error starts to go up.

$$E = \sum_{i=1}^{N} \sum_{j=1}^{p} -t_{ij} \log(y_{ij}) \tag{4}$$

Here, $t_{ij}$ is the goal of the *i-th* learning sample, which is either 1 or 0 depending on whether the *i-th* learning sample belongs to the *j-th* class or not, and $y_{ij}$ is the chance that the *i-th* learning sample belongs to the *j-th* class. The number of the training sample is $N$, and the number of different classes is $p$.

### 3.3.3 Ensemble

To obtain a good hypothesis that would produce accurate predictions for a specific issue, supervised learning algorithms explore through a hypothesis space. It could be exceedingly challenging to identify a suitable hypothesis, even if the hypothesis space contains ones that are very well suited for a specific situation. A more accurate hypothesis is created by combining several different hypotheses in an ensemble. Usually, approaches that produce many hypotheses while utilizing a single base learner are referred to as "ensembles."

A classification method that ensembles all of the hypotheses in the hypothesis space is the Bayes optimum classifier. No other ensemble can, on average, outperform it. A variation of this that makes the calculation more practical is the naive Bayes optimum classifier, which assumes that the data is conditionally independent of the class. Each hypothesis receives a vote based on how likely it is that the training dataset will be drawn from a system if it is correct. The vote of each hypothesis is additionally multiplied by the prior probability of that hypothesis to enable training data of finite size. Figure 4 shows the block diagram of the Ensemble.

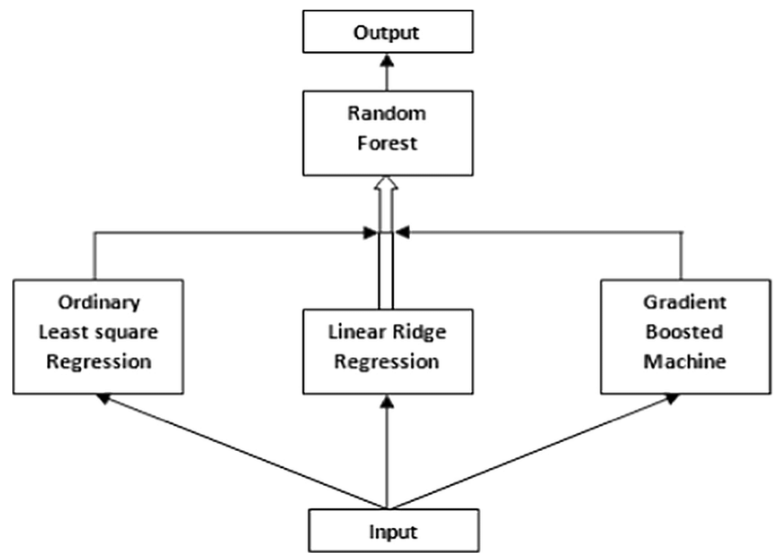

Figure 4: Block diagram of Ensemble

3.3.4 KNN

A non-parametric supervised learning technique used for classification and regression is the KNN algorithm, often known as the k-nearest neighbor algorithm. The k closest training examples in a data set serve as the input in both scenarios, and the output depends on whether KNN is being used for classification or regression.

- KNN classification - a class that an object is allocated to based on the majority vote of its k closest neighbors is determined by the item's neighbors (k is a typically small positive integer). The item is simply put into the class of its one nearest neighbor if k = 1.
- KNN regression - it is the object's property value. The average of the values of the k closest neighbors makes up this number.

An effective method for both regression and classification is to give neighbors' contributions weights, with the closest neighbors contributing more to the average than the farther neighbors. A typical weighting method, for instance, assigns each neighbors a weight of 1/d, where d is the distance between the neighbors.

When using KNN classification or regression, the neighbors are chosen from a collection of objects for which the class or object property value is known. Although there is no need for an explicit training phase, this may be considered as the algorithm's training set.

*3.4 Proposed Method*

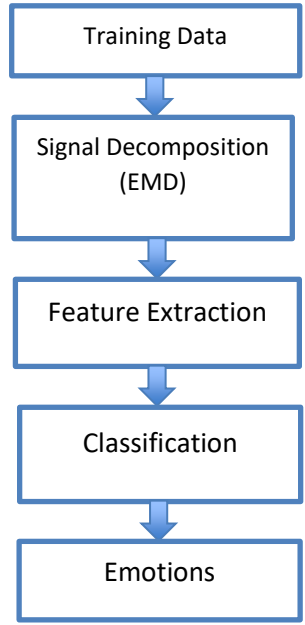

Figure 5: The flowchart of the proposed method

The flowchart of the proposed method is shown in Figure 5. At first, we load two datasets, namely, CREMA-D and TESS. The emotion class feature of those two data sets is merged together. Then we decompose the signal using the EMD. After that, we extract 14 distinct features which contain 66 sub-features. Then we train data using the SVM, KNN, Ensemble, and NNN classifiers and get different emotions (i.e., class values such as happiness, sadness, etc.). At last, we apply the classifier to the test data set and measure the accuracy of the classifiers.

## 4. DATASET

TESS Dataset: 200 target phrases are set for two actresses (aged 26 and 64) and their voice is recorded. Then the resulting recordings are categorized into seven emotions that they conveyed (i.e., anger, disgust, fear, happiness, pleasant surprise, sadness, and neutral) [15]. There are 2800 samples in those audio files for each of the two actresses. The format of the audio file is WAV.

CREMA-D Dataset: It is a collection of audio-visual data for emotion recognition [13]. The collection of data consists of visual and verbal manifestations of six fundamental emotional states, namely, anger, disgust, fear, happiness, neutral, and sad. The sample of the six emotional states is shown in Figure 6. 7442 footage is obtained, with 91 actors from various racial and ethnic backgrounds.

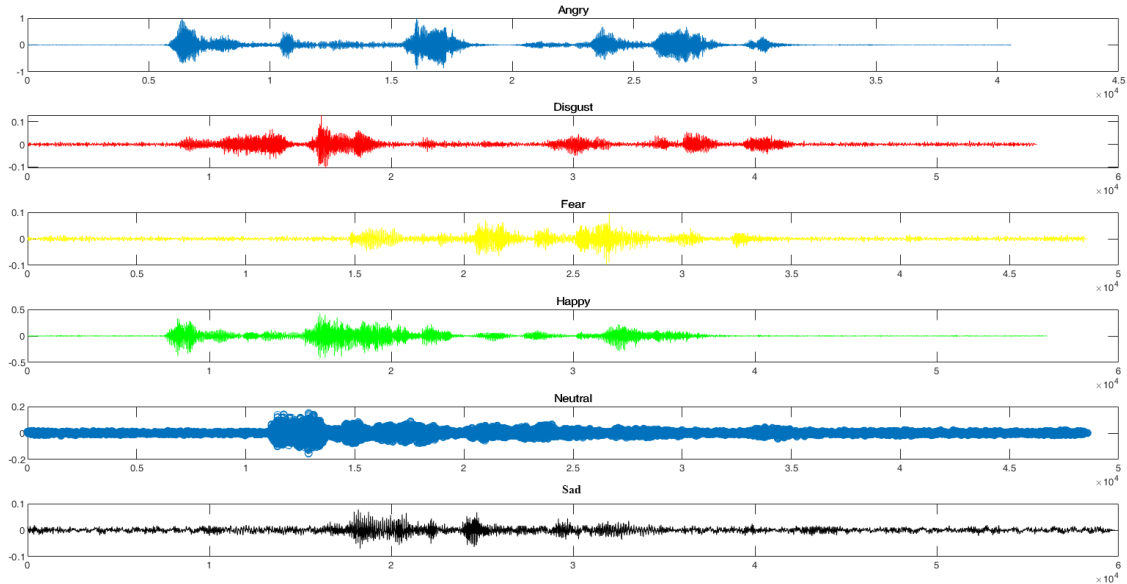

Figure 6: Data sample of CREMA-D Dataset

We have merged the above two databases and get a total of 10,242 samples; angry (1668), disgust (1664), fear (1687), happy (1663), neutral (1479), pleasant surprise (396), and sad (1663). There are some prominent characteristics of the dataset are as follows,

- CREMA-D and TESS are data set that contains 10,242 samples, with 93 different actors.
- 2-3 Second Average
- These excerpts were provided by a total of 48 male and 45 female performers ranging in age from 20 to 74 and from a diverse range of racial and ethnic backgrounds.
- Varieties of Ethnicity (African American, Canadian, Asian, Caucasian, Hispanic, and Unspecified).
- A limited number of speakers may be utilized in the production of certain sound datasets, resulting in significant information leakage.
- Emotion Class- Anger, Fear, Happy, Disgust, Neutral, Pleasant Surprise, and Sad
- Emotion Levels- Low, High, Medium, and Unspecified
- 16 kHz Sampling Rate

## 5. RESULT AND DISCUSSIONS

*5.1 Result Analysis*

In the combined databases, the total sample size is 10242 where 80% data are for training and 20% data are for testing. We apply several ML algorithms on the test data where the accuracy of SVM is 67.7%, NN is 63.3%, Ensemble is 61.6%, and KNN is 59.1% (See Table 2). However, in the training data, we get much higher accuracy of the algorithms which are 77.7 %, 76.1%, 99.1%, and 61.2% respectively for the SVM, NN, Ensemble, and KNN (See Table 3).

Table 2: The result of the test set

| Dataset | Emotions | Training Sample | Test Sample | ACCURACY | | | |
|---|---|---|---|---|---|---|---|
| | | | | SVM | NN | ENSEMBLE | KNN |
| CREMA-D | 6 | 5938 | 1488 | 61.01% | 58.22% | 55.29% | 48.7% |
| TESS | 7 | 2,240 | 560 | 83.31% | 84.5% | 77.6% | 83.3% |
| CREMA+TESS Combine | 7 | 8,194 | 2048 | 67.7% | 63.3% | 61.6% | 59.0% |

Table 3: The result of the training set

| Dataset | Emotions | Sample Size | ACCURACY | | | |
|---|---|---|---|---|---|---|
| | | | SVM | NN | ENSEMBLE | KNN |
| CREMA-D | 6 | 7,422 | 73.1% | 70.06 | 99.90% | 48.7 |
| TESS | 7 | 2,800 | 99.1% | 92.5 | 100% | 89 |
| CREMA+TESS | 7 | 10,242 | 77.7% | 76.1% | 99.1% | 61.2% |

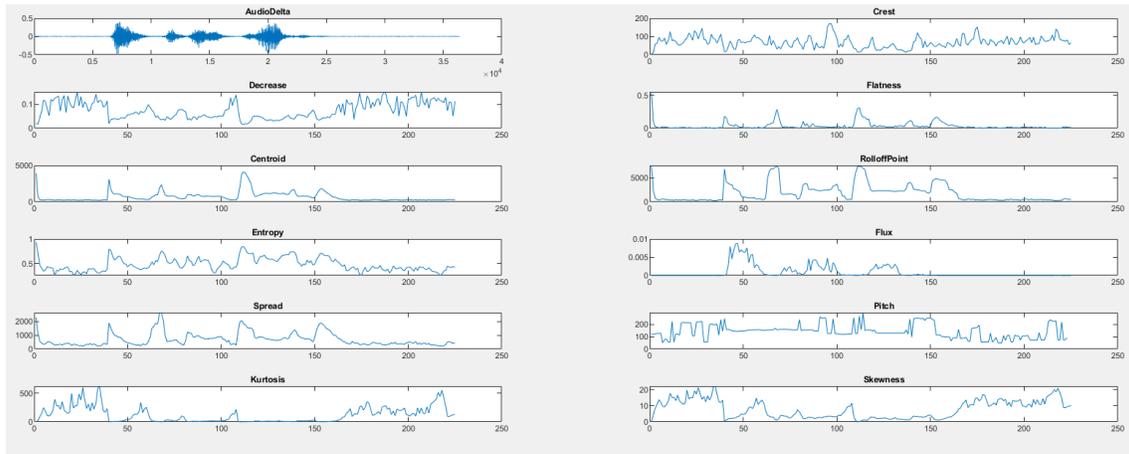

Figure 7: Feature extraction plot diagram for a single audio sample

We try different procedures/techniques to attain a better result. We observe a considerable improvement in accuracy when we train data for fewer emotional output classes. We leverage 17 unique feature extraction techniques that are used to create 66 features with their mean and variance. Figure 7 shows a plot diagram for the different features of a single audio sample. Figure 8 shows a scatter diagram of the combined dataset (CREMA and TESS) for the centroid variance and mean. Figure 9 shows the confusion matrix for the SVM algorithm.

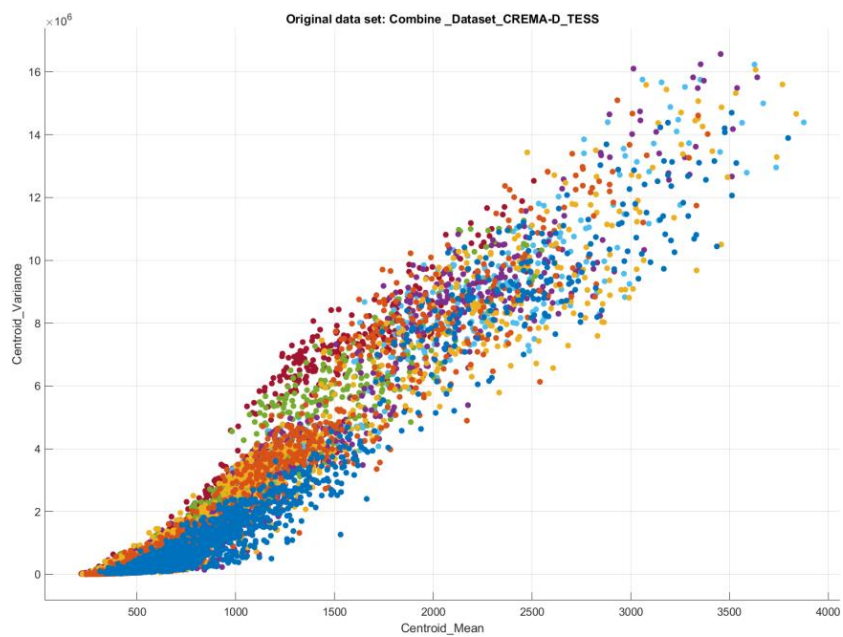

Figure 8: The diagram of the centroid variance and mean

Figure 9: Confusion Matrix for SVM algorithm

*5.2 Discussion*

Emotion recognition is a technique used in call centers to classify calls according to the feelings conveyed in them. Emotion recognition is used as a performance measure for conversational analysis, allowing for the identification of dissatisfied customers, customer contentment, and so forth. In addition, such techniques might be used to observe how companies engage with their customers through the use of call centers. It is necessary to incorporate a human specialist with restricted capabilities to analyze emotions during such interactions at that moment. However, if one uses machines to complete the operation, the cost would be far lower and the output would be more consistent.

**Advantages:**

- It helps employees and HR (Human Resource) teams manage stress.
- Real-time vocal emotion analysis allows automated customer support representatives to recognize customers' emotional states.
- This lets organizations build emotional ties with customers via virtual assistants.
- This technology helps caregivers or other family members (by alerting) to care for the babies and the elder persons in the house.
- By analyzing emotions, social media can classify sensitive and important information.

**Disadvantages:**

- Validation of the emotion datasets is difficult to achieve accurate emotion recognition.
- Emotion detection is a haystack process.
- Determination/classification of emotions in various languages and accents is very difficult.
- Some software only accepts text data where image, pattern, video, and audio inputs are invalid.

## 6. CONCLUSION

A technique for emotion recognition in speech using SVM, NN, Ensemble, and KNN classifiers is presented in this paper. We have classified seven emotions namely, anger, disgust, fear, happiness, neutral, pleasant surprise, and sad in the CREMA–D and TESS data set. EMD is used to decompose the signal and MFCC, GTCC, and Audio Spectral for emotion prediction. We have extracted 19 distinct features which contain 66 sub-features. Data Augmentation is applied to the audio samples to increase the variety of datasets thereby improving the model's performance. The SVM has the highest accuracy among the four ML algorithms. In the future, we may select fewer features and more emotion classes and try to achieve higher accuracy.

## Declaration of Competing Interests

The authors declare that they have no known competing financial interests or personal relationships that could have appeared to influence the work reported in this paper.

## Authors Biography

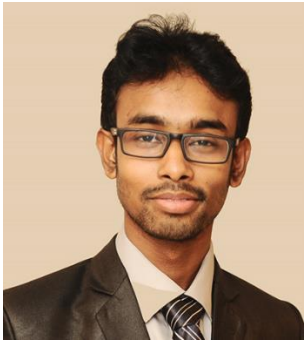

**MEHRAB HOSAIN** is a graduate student at Delhi Technological University in India, where he is pursuing a Master of Technology in Signal Processing and Digital Design. He previously earned an M.Sc. in Computer Science and Engineering from the University of Information Technology and Sciences (UITS) in Bangladesh and a B.Sc. in Electrical and Electronic Engineering from the Bangladesh University of Business and Technology (BUBT). He has five years of experience as an IT professional at Impress Telefilm Limited (Channel i) in Bangladesh and has been a highly-rated freelancer on Upwork for seven years. His research interests include signal processing, computer vision, machine learning, artificial intelligence, data science, and digital forensics.

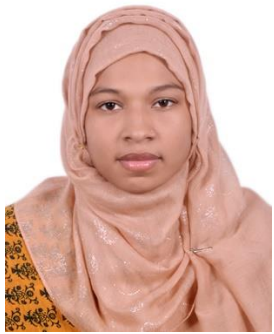

**MOST. YEASMIN ARAFAT** is an ambitious and highly skilled individual who is currently pursuing a Master's degree in Computer Science and Engineering at the University of Information Technology and Sciences in Bangladesh. She has already earned a Bachelor of Science in Electrical and Electronic Engineering from the Bangladesh University of Business and Technology and a Diploma in Bio-Medical Engineering from the Dhaka Mohila Polytechnic Institute. She has two years of professional experience as an Assistant Engineer at Icon Engineering Services Limited. Her areas of expertise are Computer Vision, Data Analytics, Image Processing, Machine Learning, and Artificial Intelligence.

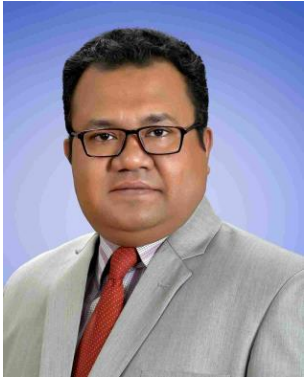

**GAZI ZAHIRUL ISLAM** obtained his Ph.D. at the Bangladesh University of Professionals. He completed M.Sc. in wireless communications systems engineering from the University of Greenwich, UK, and a B.Sc. in computer science and engineering from Chittagong University of Engineering and Technology (CUET), Bangladesh. He is currently teaching as an Associate Professor at Southeast University, Bangladesh. He had published a good number of research papers in renowned journals and conferences. The author reviewed many reputed journals such as IET Communications, IEEE Access, Indonesian Journal of Electrical Engineering and Computer Science, Journal of King Saud University - Computer and Information Sciences, Journal of Communications, Current Research in Behavioral Sciences, Internet of Things and Cyber-Physical Systems, etc. He conducted research and several projects on Wi-Fi networks, wireless communications, computer networking, Internet of Things, software/app development, e-governance/e-commerce, Machine Learning, etc.

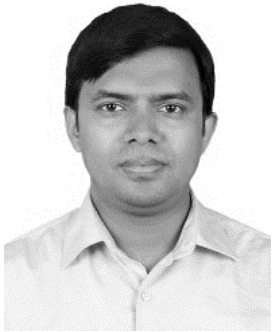

**JIA UDDIN** received Ph.D. in Computer Engineering from the University of Ulsan, Korea, in January 2015. He is an Assistant Professor in AI and Big Data Department, Endicott College, Woosong University, South Korea and an Associate Professor (On Leave), Computer Science and Engineering Department at BRAC University, Bangladesh. His research interests include fault diagnosis, computer vision, and multimedia signal processing.

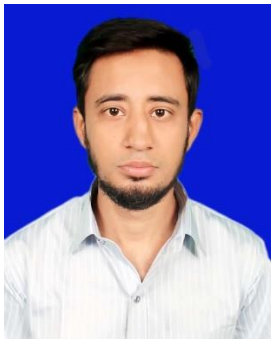

**MD. MOBARAK HOSSAIN** is currently working as a Computer Network & System Administrator. He completed B.Sc. in Computer Science and Engineering from the Daffodil International University, Bangladesh in 2021. He has good experience in the Internet of Things (IoT), Computer Networking, Cloud Service, Windows Server, Cyber Security. He has been working as a System Support Engineer at Mutual Trust Bank, Bangladesh. His interested working areas are Network & System Administration, Security Specialization, Cyber Security, and Cloud Computing. His research interest is in the Internet of Things (IoT), Artificial Intelligence, Computer Networks, and Cyber Security.

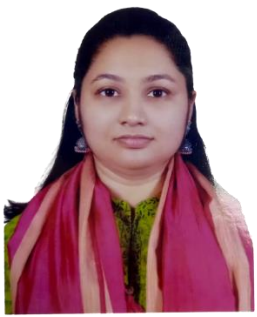

**FATEMA ALAM** graduated from the Bangladesh University of Business and Technology (BUBT) with a Bachelor's degree in Electrical and Electronics Engineering. During her Bachelor's degree, she completed an academic research project on "SPEAKER INDEPENDENT SPEECH RECOGNITION USING DEEP FEED-FORWARD NEURAL NETWORK" due to her interest in Machine Learning, Artificial Intelligence, and Data Science. She is presently pursuing a Master's degree in Applied Physics and Electronics at Jahangirnagar University in Bangladesh. Her research interests include the area of Applied Physics and Digital Electronics along with Data Analytics, Machine Learning, and Artificial Intelligence.